\documentclass[journal]{IEEEtran}

\usepackage{cite}

\ifCLASSINFOpdf
   \usepackage[pdftex]{graphicx}
   \graphicspath{{../pdf/}{../jpeg/}}
   \DeclareGraphicsExtensions{.pdf,.jpeg,.png}
\else
\fi
\UseRawInputEncoding
\usepackage{amsmath}
\usepackage{array}
\usepackage{standalone}
\usepackage{etoolbox}
\newtoggle{arXiv}
\newtoggle{bibrun}
\usepackage{newtxtext, newtxmath}
\usepackage{graphicx}
\usepackage{color, calc}
\usepackage{array, booktabs}
\usepackage{tikz, tikz-3dplot}
\usetikzlibrary{arrows, arrows.meta, calc, positioning, decorations.pathmorphing}
\tikzset{>=Stealth}
\usepackage{siunitx, physics}

\usepackage{enumitem}
\setlist[description]{labelindent=0pt, leftmargin=\parindent, font=\normalfont\itshape}
\usepackage{url}
\usepackage{subfigure}
\usepackage{textcomp}
\usepackage{eurosym}
\usepackage{xfrac}
\usepackage{pgfplots}
\usepackage{stmaryrd}
\pgfplotsset{compat=1.17}
\usepackage{makecell}
\usepackage[utf8]{inputenc}

\newcommand{\m}{MaRCoS}
\newcommand{\mg}{MaRGE}

\begin{document}

\title{\textit{In-vivo} imaging with a low-cost MRI scanner and cloud data processing in low-resource settings}

\author{\IEEEauthorblockN{
Teresa~Guallart-Naval\IEEEauthorrefmark{1},
Robert~Asiimwe\IEEEauthorrefmark{2},
Patricia~Tusiime\IEEEauthorrefmark{2},
Mary~A.~Nassejje\IEEEauthorrefmark{2},
Leo~Kinyera\IEEEauthorrefmark{2},
Lemi~Robin\IEEEauthorrefmark{2},
Maureen~Nayebare\IEEEauthorrefmark{2}\IEEEauthorrefmark{7},
Luiz~G.~C.~Santos\IEEEauthorrefmark{1},
Marina~Fern\'andez-Garc\'ia\IEEEauthorrefmark{1},
Lucas~Swistunow\IEEEauthorrefmark{1},
Jos\'e~M.~Algar\'in\IEEEauthorrefmark{1},
John~Stairs\IEEEauthorrefmark{4},
Michael~Hansen\IEEEauthorrefmark{4},
Ronald~Amodoi\IEEEauthorrefmark{2},
Andrew~Webb\IEEEauthorrefmark{3},
Joshua~Harper\IEEEauthorrefmark{5},
Steven~J.~Schiff\IEEEauthorrefmark{6},
Johnes~Obungoloch\IEEEauthorrefmark{2},
and~Joseba~Alonso\IEEEauthorrefmark{1}}

\IEEEauthorblockA{\IEEEauthorrefmark{1}MRILab, Institute for Molecular Imaging and Instrumentation (i3M), Consejo Superior de Investigaciones Cient\'ificas (CSIC) \& Universitat Polit\`ecnica de Val\`encia (UPV), Valencia, Spain}\\
\IEEEauthorblockA{\IEEEauthorrefmark{2}Mbarara University of Science and Technology (MUST), Dept. of Biomedical Engineering, Mbarara, Uganda}\\
\IEEEauthorblockA{\IEEEauthorrefmark{7}McGill University, department of Biological and Biomedical Engineering, Montreal, Canada}\\
\IEEEauthorblockA{\IEEEauthorrefmark{3}Leiden University Medical Center (LUMC), Dept. of Radiology, Leiden, Netherlands}\\
\IEEEauthorblockA{\IEEEauthorrefmark{4}Microsoft Research, Health Futures, Washington, USA}\\
\IEEEauthorblockA{\IEEEauthorrefmark{5}Facultad de Ciencias Aplicadas, Universidad Comunera, Asuncion, Paraguay}\\
\IEEEauthorblockA{\IEEEauthorrefmark{6}Yale University, Dept. of Neurosurgery and Dept. of Epidemiology of Microbial Diseases, New Haven, USA}

\thanks{Corresponding author: J. Alonso (joseba.alonso@i3m.upv.es).}
}


\maketitle

\begin{abstract}
\newline
Purpose: {\normalfont To demonstrate \textit{in-vivo} imaging with a low-cost, low-field MRI scanner built and operated in Africa, and to show how systematic hardware and software improvements can mitigate the main operational limitations encountered in low-resource environments.}\\
Methods: {\normalfont A 46\,mT Halbach scanner located at the Mbarara University of Science and Technology (Uganda) was upgraded through a complete reorganization of grounding and shielding, installation of new control electronics and open-source user-interface software. Noise performance was quantified using a standardized protocol and \textit{in-vivo} brain images were acquired with three-dimensional RARE sequences. Distortion correction was implemented using cloud-based reconstructions incorporating magnetic field maps.}\\
Results: {\normalfont The revamped system reached noise levels routinely below three times the thermal limit and demonstrated stable operation over multi-day measurements. Three-dimensional T$_1$- and T$_2$-weighted brain images were successfully acquired and distortion-corrected with remote GPU-based reconstructions and near real-time visualization through the user interface.}\\
Conclusions: {\normalfont The results show that low-cost MRI systems can achieve clinically relevant image quality when electromagnetic noise and power-grid instabilities are properly addressed. This work highlights the feasibility of sustainable MRI development in low-resource settings and identifies stable power delivery and local capacity building as the key next steps toward clinical translation.}
\end{abstract}

\IEEEpeerreviewmaketitle

\section{Introduction}
\IEEEPARstart{T}{he} lack of access to essential radiological infrastructure compromises the quality of healthcare services and signifies a prominent burden on Low and Middle Income Countries (LMICs). Specifically, present-day Magnetic Resonance Imaging (MRI) technologies remain out of reach for about 70\,\% of the world population \cite{Murali2024,Geethanath2019}, despite estimates that millions of lives could be saved every year by making it more widely available \cite{Hricak2021}.

The MRI accessibility problem takes different shapes in different regions \cite{Murali2024,Geethanath2019}. In Sub-Saharan Africa (SSA), formed by 54 countries and with a total population of over 1{.}2 billion people \cite{SSAPop}, a recent needs assessment survey identified brain, spine and musculoskeletal imaging applications as most relevant \cite{Anazodo2023}. This is motivated by a number of factors. One is that SSA accounts for roughly half of the global incidence of hydrocephalus cases (90\,\% of the cases appear in LMICs, mostly in pediatric populations) \cite{dewan2018}. While ultrasound and CT are commonly used for initial diagnosis, MRI plays a key role in non-ionizing assessment, surgical planning, and post-shunting monitoring \cite{harper2021}. Another is that the burden of non-communicable diseases in Africa steadily increases as environmental and societal trends evolve. For instance, stroke is now more prevalent in the African continent than in Western Europe and the USA \cite{Anazodo2023}, and MRI is the recommended standard of care for these (and multiple other) neurological diseases \cite{Bhat2021}. Spine and musculoskeletal applications are motivated mostly by traffic accidents. For reference, traffic deaths on African roads add up to about one quarter of the victims worldwide, while possessing barely 2\,\% of the world's vehicle fleet \cite{Carnis2023,RoadSafety}. Likewise, the incidence of non-fatal accidents is extreme in SSA, and MRI could play a key role for traumatic spine \cite{Kumar2016} and extremity injuries \cite{Deutsch1989,Deyle2011}.

However, MRI machines are complex, proprietary and extraordinarily expensive. Furthermore, they necessitate highly-trained staff for their operation. These aspects, combined with the scarce economic, material, and trained human resources available, preclude the penetration of present-day MRI in SSA and other LMICs. Low-field MRI technologies (LF-MRI) operate below 0.5\,T, use permanent rather than superconducting magnets, and have been postulated as a means to alleviate this situation \cite{Sarracanie2015,Marques2019,Sarracanie2020,Wald2020,Bhat2021,harper2021,Webb2023}. Relevant milestones achieved with LF-MRI systems include point-of-care imaging \cite{McDaniel2019,Nakagomi2019,Cooley2020,OReilly2020,Sheth2021,Mazurek2021,Liu2021}, home healthcare \cite{Guallart-Naval2022}, quantitative MRI and fingerprinting \cite{OReilly2021,Sarracanie2021}, hard-tissue imaging \cite{Algarin2020,Borreguero2025,Borreguero2025b}, artifact-free imaging of metallic implants \cite{Guallart-Naval2022,VanSpeybroeck2021}, robust 3D and diffusion imaging at low field \cite{Qiu2024}, or whole-body imaging \cite{Zhao2024}. Still, as of today, translation to the clinic of most of the prospective LF-MRI applications remains to be demonstrated, more so for deployment in low-resource settings, where there are multiple operational challenges other than the system cost (see Sec.~\ref{sec:baseline}).

In this paper we present \emph{in-vivo} images taken with a low-cost ($\sim\SI{30}{k\$}$) LF-MRI system built in Africa. The 46\,mT Halbach scanner had already been constructed on site at the Mbarara University of Science and Technology (MUST), in Uganda \cite{Obungoloch2023}. To make it suitable for \emph{in-vivo} imaging, we have revised the complete electronics setup and installed the latest stable versions of the control electronics and user interface applications (\m{} and \mg{}, \cite{Negnevitsky2023,Algarin2024}). Additionally, we have upgraded \m{} to integrate with Tyger, an open-source framework for remote signal processing \cite{Tyger,Guallart-Naval2025b}. As part of this work, the MUST team has gathered and documented know-how and methodologies followed in the laboratories of the Institute for Molecular Imaging and Instrumentation (i3M, Valencia, Spain) and the Leiden University Medical Center (LUMC, Leiden, The Netherlands), thereby markedly increasing the local expertise and autonomy.

\section{System Baseline and Challenges}
\label{sec:baseline}

\subsection{System baseline}

\begin{figure*}[t]
  \centering
  \includegraphics[width=\textwidth]{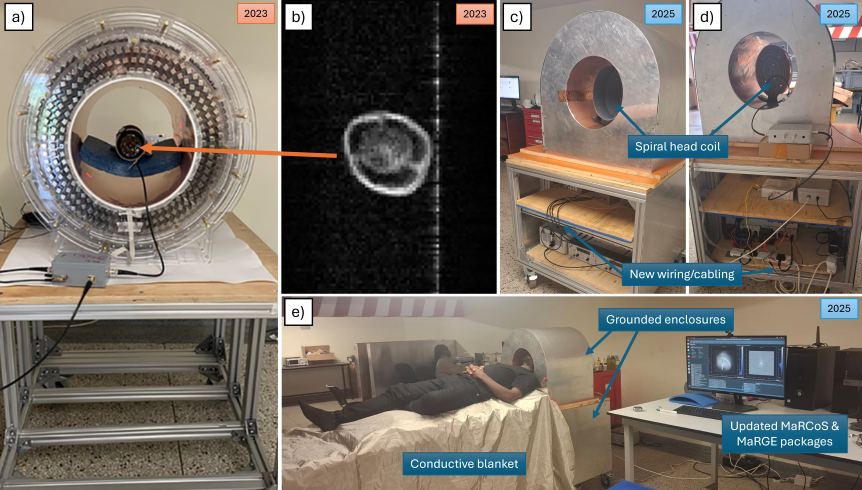}
  \caption{System evolution at MUST. a) Scanner after the initial assembly in 2023. b) First reconstructed image (bell pepper). Front (c), rear (d), and global (e) views of the MUST scanner in the present configuration, highlighting the improved electronics setup, RF coils, and mechanical integration.}
  \label{fig:setup}
\end{figure*}

The MRI scanner at MUST \cite{Obungoloch2023} is built around a \SI{46}{mT} permanent magnet array in a Halbach configuration, without any passive shimming. Over a 20\,cm diameter spherical volume, the resulting $B_0$ inhomogeneity is approximately 4,700\,ppm. The magnet weighs $<100$\,kg and is mounted on an aluminum trolley with supports for the gradient and radio-frequency (RF) subsystems. The gradient set was manufactured in Uganda from copper wire wound onto 3D-printed formers, following designs supplied by the collaborating institutions. The RF chain consisted of a single solenoidal coil connected to a non-gated, commercial low-noise amplifier (LNA) and a home-made unblanked transmit/receive (TxRx) switch. Power amplification for the gradients relied on an open-source design \cite{OSII} and RF transmission made use of a commercial 1\,W power amplifier (RFPA). The scanner was controlled using an early version of the open-source \m{} platform. At the time, the software did not include the \mg{} graphical interface, and all operation was handled via low-level scripts and command-line tools.

This installation at MUST in 2023 represented the first documented on-site construction of an MRI system in SSA, enabling basic imaging (Fig.~\ref{fig:setup} a)-b)) and marking an important proof of concept that demonstrated the feasibility of low-field MRI in a resource-limited environment.

While the system produced images, its operation revealed important limitations for progression toward \emph{in-vivo} imaging. Most prominently, the early images suffered from low SNR and geometric distortions, falling short of the robustness and usability required for clinical translation. For reference, the bell pepper image in Figure~\ref{fig:setup} took hours to acquire. 

\subsection{Challenges}

Here we summarize the operational challenges encountered for \emph{in-vivo} imaging with the 2023 setup, providing the baseline against which the hardware and software upgrades presented later in the paper should be interpreted.

\subsubsection{Electromagnetic noise and power grid reliability}

\begin{figure*}
  \centering
  \includegraphics[width=\textwidth]{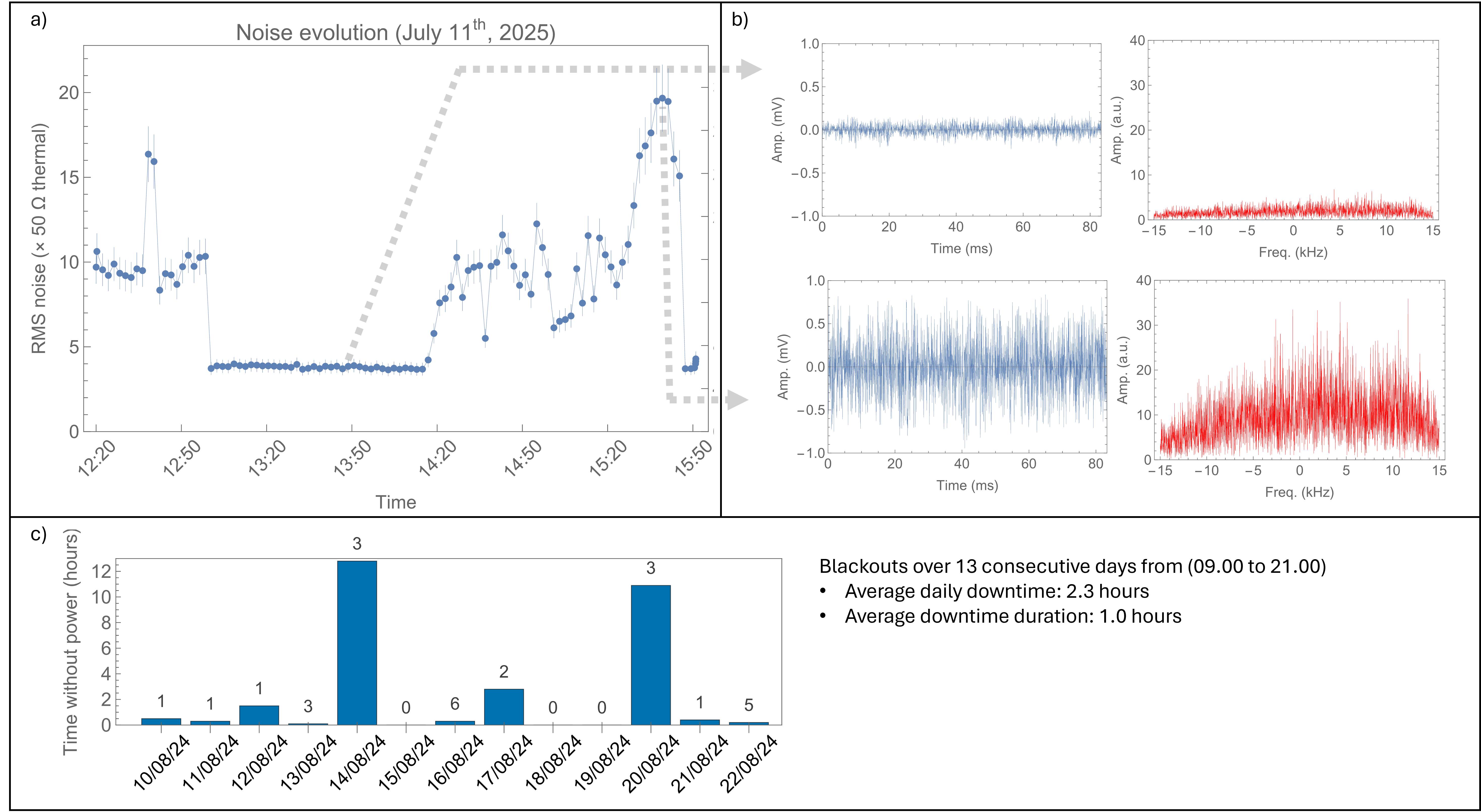}
  \caption{Electromagnetic environment at MUST. a) Noise levels logged over a few hours after the revamp but with suboptimal grounding, highlighting the strong dependence of the measured noise floor on building activity and grid conditions. b) Inserts show representative traces and spectra under ``quiet'' and ``noisy'' conditions. c) Frequency of power-off events recorded over a two-week period.}
  \label{fig:long_noise_mmt}
\end{figure*}

Operation at low magnetic field strengths is intrinsically challenging because the induced NMR signals are weak and image quality is ultimately limited by the signal-to-noise ratio (SNR) \cite{Guallart-Naval2025a}. In the baseline MUST system this problem was exacerbated by the absence of any RF shielding around the magnet and, more importantly, by suboptimal grounding of the system, which left the receive chain highly exposed to radiated electromagnetic interference (EMI). As a result, the effective noise floor was sizable and fluctuated strongly even between consecutive acquisitions.

The impact of the electromagnetic environment is illustrated in Figure~\ref{fig:long_noise_mmt}a)-b), which show noise levels logged over a few hours. Even with the later, optimized version of the scanner, the measurements reveal a clear dependence on electrical activity within the building and on the state of the local power grid, if grounding is suboptimal.

On top of this already noisy environment, the reliability of the grid itself represents an additional limitation. In Mbarara, extended outages and short blackouts are frequent, forcing scanner downtime and occasionally corrupting acquisitions, despite the diesel generator available in campus. Figure~\ref{fig:long_noise_mmt}c) summarizes the power interruptions recorded over a two-week period, underscoring the need for autonomous and stable power solutions.

\subsubsection{Access to material and computational resources}
Carrying out MRI research and system development in low-resource settings is inevitably constrained by the limited availability of technical components and laboratory infrastructure. In the case of the MUST scanner, access to basic electronic parts such as connectors, coaxial cables, or measurement instruments is limited, often requiring long procurement times or international shipping. These limitations compound the effects of electromagnetic interference discussed above, since effective noise suppression and shielding depend critically on appropriate materials and hardware.

In this regard, illustrative examples encountered during this work include: i) the fact that the system operates with a \SI{1}{W} RF power amplifier, whereas similar setups in partner laboratories employ units in the \SIrange{250}{1,000}{W} range; ii) when one of the diodes in the passive TxRx switch failed, its replacement was delayed by the absence of local suppliers, and standard diodes that cost only a few cents elsewhere had to be substituted with a lower-grade alternative that was both more expensive and slow to obtain; iii) precision mechanical work is also challenging and the lack of laser-cutting services in Mbarara limited the ability to fabricate accurate shimming inserts, hindering fine adjustment of the magnetic field homogeneity as routinely performed in comparable Halbach magnets \cite{OReilly2022}.

Finally, computational resources present an additional bottleneck. High-end workstations and GPUs are not readily available, and subscription-based cloud services are often unaffordable. Even the control computer used to operate the MUST scanner offers modest performance, affecting data throughput and the ability to perform advanced reconstructions locally.

\section{Methods}

The two main activities carried out in this work were the upgrade of the MUST scanner to achieve improved operational capabilities and the acquisition of \emph{in-vivo} images to demonstrate the impact of these improvements.

\subsection{Scanner revamp and electromagnetic noise suppression}
\label{sec:met_noise}

The scanner revamp focused on four main aspects (Fig.~\ref{fig:setup}c)-e)): a complete overhaul of grounding and shielding; multiple new RF coils and impedance-matching electronics; the installation of the latest releases of \m{} and \mg{}; and the integration of Tyger for cloud-based data processing and image reconstruction.

\begin{figure}
  \centering
  \includegraphics[width=0.85\columnwidth]{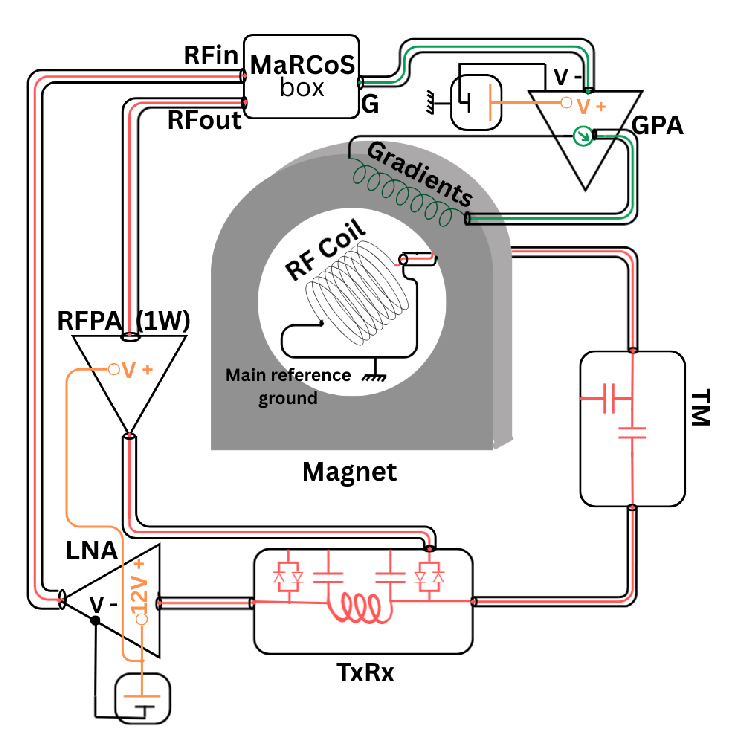}
  \caption{Schematic of the complete electronics setup of the MUST scanner after the revamp. 
    Red lines indicate RF signal paths (transmit and receive), green lines correspond to gradient drive signals, orange lines denote power connections, and black lines represent ground. The diagram highlights the main subsystems, including the magnet, RF and gradient coils, TxRx switch, LNA, 1\,W RFPA, GPA, and the \m{} control unit.}    
  \label{fig:gnd_scheme}
\end{figure}

A comprehensive reorganization of the grounding and shielding layout was carried out following the guidelines detailed in Ref.~\cite{Guallart-Naval2025a}. Particular attention was paid to cable management, enclosure integrity, and the definition of a single, low-impedance ground reference for the entire system (Fig.~\ref{fig:gnd_scheme}). All electronic subsystems were enclosed in conductive aluminum housings with robust electrical contact between panels and lids, ensuring continuous shielding and minimizing radiative leakage. Internal RF shielding around the receive coil was reinforced with a short, thick connection between the coil return and the inner copper sleeve, while the outer shield was bonded to all major enclosures and cable shields following, where possible, a star-grounding configuration. Gradient cables were rerouted and shielded to suppress inductive coupling, and all ``noisy'' elements (such as power supplies, digital control boards, and switching regulators) were relocated away from the RF path. Only linear power supplies were used for the analog subsystems, explicitly avoiding switch-mode supplies to reduce conducted and radiated EMI.

In addition, any component containing a ferromagnetic core and directly driven at 50\,Hz (such as the transformers inside linear supplies, or mains-driven electromechanical relays) was either removed or positioned at a sufficient distance from the magnet. These elements generate alternating magnetic fields that modulate the Larmor frequency and introduce artifacts unless echo and repetition times are exact multiples of 20\,ms. For acquisition windows exceeding a few ms, such modulation can also distort the readout signal, as the field may vary significantly within that interval. Both effects become apparent already for sub-nano-tesla modulation depths. While increasing the distance from the magnet effectively mitigates $B_0$ modulation, it also necessitates longer cable runs, which in turn raise susceptibility to noise pickup and ground potential differences.

As part of the hardware upgrade, several new RF coils were constructed to replace the single cylindrical solenoid used in the original setup (Fig.~\ref{fig:setup}c)-d)). The primary head coil consists of a single-layer spiral geometry adapted to the head anatomy, following the design described in Ref.~\cite{Sarracanie2020}. This provides an improved filling factor and enhanced SNR for brain imaging. In addition, two solenoidal coils were built: one optimized for head imaging and another, smaller version for e.g. knee acquisitions. A new impedance tuning and matching (TM) box was also developed, incorporating multiple capacitors on rotary switches that allow manual selection among discrete capacitance values without the need for soldering.

The control and user interface software were upgraded to the latest stable releases of \m{} and \mg{}~\cite{Algarin2024}. The new \mg{} version provides a unified graphical environment programmed in Python, integrating pulse sequence control, system calibration, and reconstruction tools within a modular layout. It replaces the early command-line interface used in the original system, offering enhanced usability, real-time monitoring, and compatibility with standardized sequence formats such as Pulseq.

The system was also equipped with Tyger~\cite{Guallart-Naval2025b}, an open-source platform that enables streaming of raw MR data from the scanner to cloud-based computational resources, where reconstructions or post-processing tasks can be executed inside Docker containers.

\begin{figure}
  \centering
  \includegraphics[width=\columnwidth]{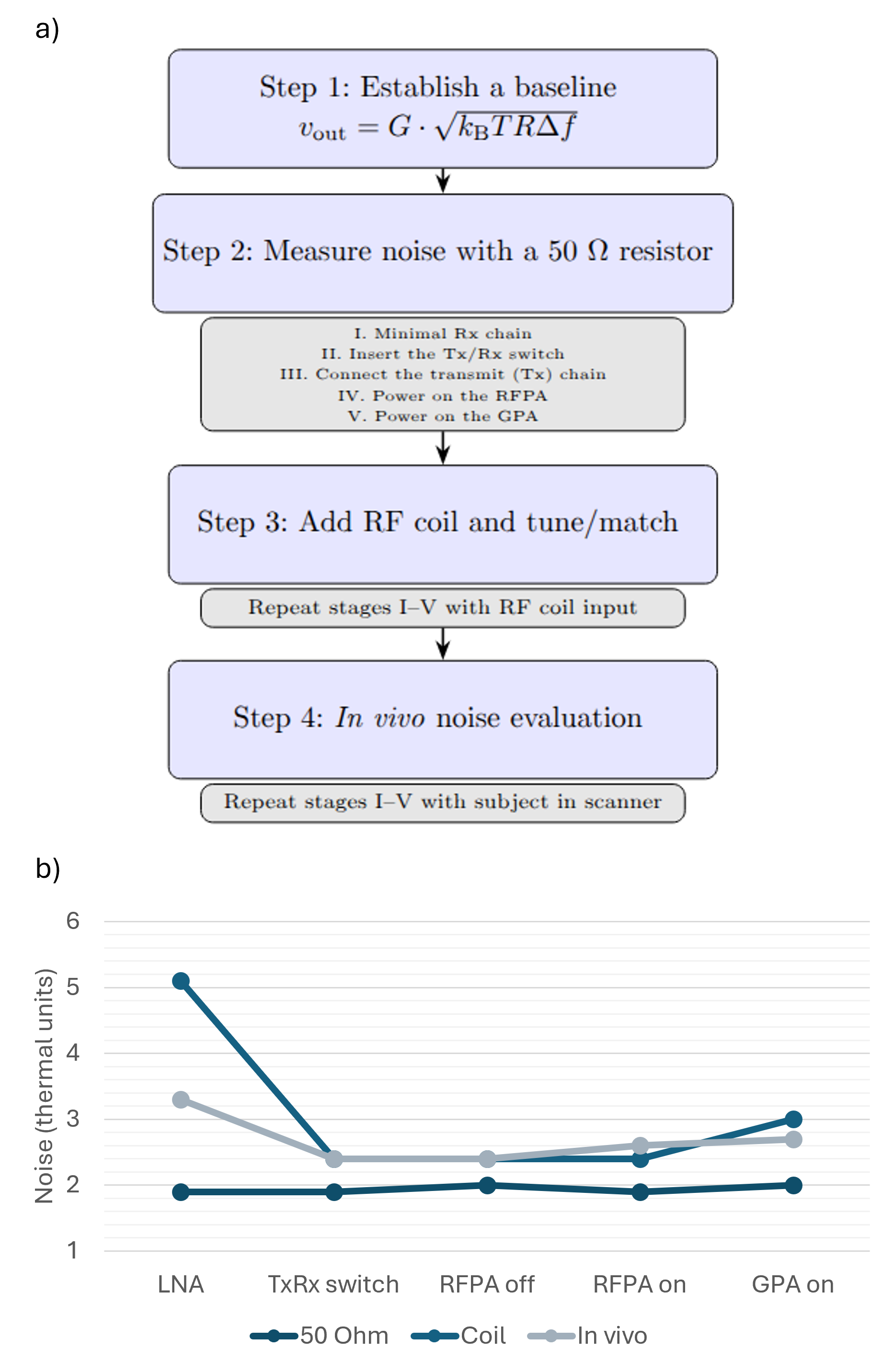}
    \caption{Electromagnetic noise characterization and suppression in the MUST scanner. a) Four-step protocol for systematic noise assessment in low-field MRI systems, adapted from Ref.~\cite{Guallart-Naval2025a}. Steps~2--4 involve repeated measurements through stages~I--V to isolate the contribution of each subsystem as the scanner is progressively assembled and powered. The expression $v_{\mathrm{out}} = G \cdot \sqrt{k_\text{B} T R \Delta f} = G \cdot v_{\mathrm{n}} / 2 $ defines the expected baseline noise voltage, where $G$ is the linear transducer gain of the amplifier, $k_\text{B} \approx \SI{1.38e-23}{J/K}$ is Boltzmann’s constant, and $v_{\mathrm{n}}$ is the RMS voltage noise generated by a resistor $R$ at temperature $T$ over a bandwidth $\Delta f$. b) Normalized noise measurements, expressed as multiples of the theoretical thermal limit, for different hardware configurations.}    
  \label{fig:protocol}
\end{figure}

Noise measurements were performed following the stepwise protocol described in Ref.~\cite{Guallart-Naval2025a}, adapted here for the MUST system (Fig.~\ref{fig:protocol}a)). In brief, we quantified the root-mean-square (RMS) voltage noise at the output of the receive chain under different configurations: first with a 50\,$\Omega$ resistor connected directly to the LNA, then with the unloaded RF coil, and finally with the coil loaded by an \emph{in-vivo} subject. Such noise measurements can be acquired with a dedicated pulse sequence in \mg{}. At MHz-range Larmor frequencies, the fundamental noise limit is set by thermal (Johnson-Nyquist) noise from the receiver impedance. For a 50\,$\Omega$ source at room temperature ($\approx 300$~K), connected to a 50\,$\Omega$ input LNA, the theoretical voltage noise spectral density is $\approx\SI{0.45}{nV/\sqrt{Hz}}$, defining a baseline that corresponds to one ``thermal unit''. Measured RMS voltages were normalized to this reference. Each measurement used a 30\,kHz bandwidth and lasted 80\,ms, covering four full cycles of the 50\,Hz mains to capture periodic interference components.

\subsection{In-vivo imaging acquisitions}
\label{sec:met_images}

\begin{table*}
\caption{Acquisition parameters for the images included in this work. All datasets were acquired using a 3D RARE sequence; TE denotes the effective echo time. Note that the image in Fig.~8-bottom corresponds to the same dataset as Fig.~6-T1w.}
\centering
\begin{tabular}{c c c c c c c c c c c c}
\toprule
Figure & \thead{FoV \\ (mm$^3$)} & \# pixels &  \thead{Partial \\ Fourier  (\%)} & \thead{Resolution \\ (mm$^3$)} & \thead{BW \\ (kHz)} & \thead{TR \\(ms)} & \thead{TE \\(ms)} & \thead{ETL} & \thead{$k$-space \\filling} & Avgs. & \thead{Scan time \\(min)} \\
\midrule
\ref{fig:T1T2}-T1w & $179\times192\times117$ & $100\times 100\times 24$  & 100 & $1.7\times1.9\times4.8$ & 16.6 & 600 & 16 & 5 & center-out  & 7 & 33 \\
\midrule
\ref{fig:T1T2}-T2w & $195\times210\times140$ & $80\times80\times18$  & 85 & $2.4\times2.6\times7.7$ & 13.3 & 2500 & 40 & 8 & linear & 7 & 44 \\
\midrule
\ref{fig:knee} & $120\times160\times120$ & $90\times 90\times 16$  & 65 & $1.7\times1.3\times7.5$ & 22.5 & 200 & 20 & 5 & center-out  & 40 & 24 \\
\midrule
\ref{fig:displaced}-top & $185\times200\times142$ & $100\times 100\times 24$  & 80 & $1.8\times2.0\times5.9$ & 16.6 & 600 & 16 & 5 & center-out  & 8 & 30 \\
\midrule
\ref{fig:displaced}-bottom & $179\times192\times117$ & $100\times 100\times 24$  & 100 & $1.7\times1.9\times4.8$ & 16.6 & 600 & 16 & 5 & center-out  & 7 & 33 \\
\midrule
\ref{fig:dist_corr} & $209\times208\times152$ & $100\times 100\times 24$  & 85 & $2.1\times2.1\times6.3$ & 16.6 & 600 & 16 & 5 & center-out  & 10 & 40 \\
\bottomrule
\end{tabular}
\label{tab:seq_params}
\end{table*}

All \emph{in-vivo} images in this work were acquired with the MUST scanner described in Sec.~\ref{sec:baseline}, after the upgrades described in Sec.~\ref{sec:met_noise}. Experiments employed three-dimensional Rapid Acquisition with Relaxation Enhancement (3D-RARE) sequences optimized for low-field operation, using a repetition time (TR) and echo spacing time (TE) suitable for either T$_1$- or T$_2$-weighted contrast. All sequence parameters are summarized in Table~\ref{tab:seq_params}.

To investigate the most favorable subject positioning within the unshimmed 46~mT magnet, further T$_1$-weighted images were acquired with the head placed at different positions inside the magnet bore. 

To mitigate the geometric distortions resulting from field inhomogeneity, we implemented a data processing pipeline that incorporates prior knowledge of the spatial $B_0$ distribution during image reconstruction. The magnetic field map was obtained using the Single-Point Double-Shot (SPDS) method~\cite{Borreguero2025}, and the reconstructed data were processed iteratively with this field information. Because these algorithms are computationally intensive, the reconstruction was executed remotely through Tyger.

\section{Results}

\subsection{Scanner revamp and electromagnetic noise suppression}
\label{sec:res_noise}


Figure~\ref{fig:protocol}b) summarizes the normalized RMS noise measurements obtained under the three configurations introduced in Sec.~\ref{sec:met_noise}. The table follows the structure of Ref.~\cite{Guallart-Naval2025a}, where each row corresponds to a specific hardware configuration. Values are expressed in units of the thermal limit. The first line (``LNA'') corresponds to a minimal receive chain comprising only the low-noise amplifier and digitizer, with all other components physically disconnected. The subsequent rows show the progressive addition of subsystems: insertion of the TxRx switch, connection of the transmit electronics, activation of the RFPA, and activation of the gradient power amplifiers (GPA).

\begin{figure*}[t]
  \centering
  \includegraphics[width=\textwidth]{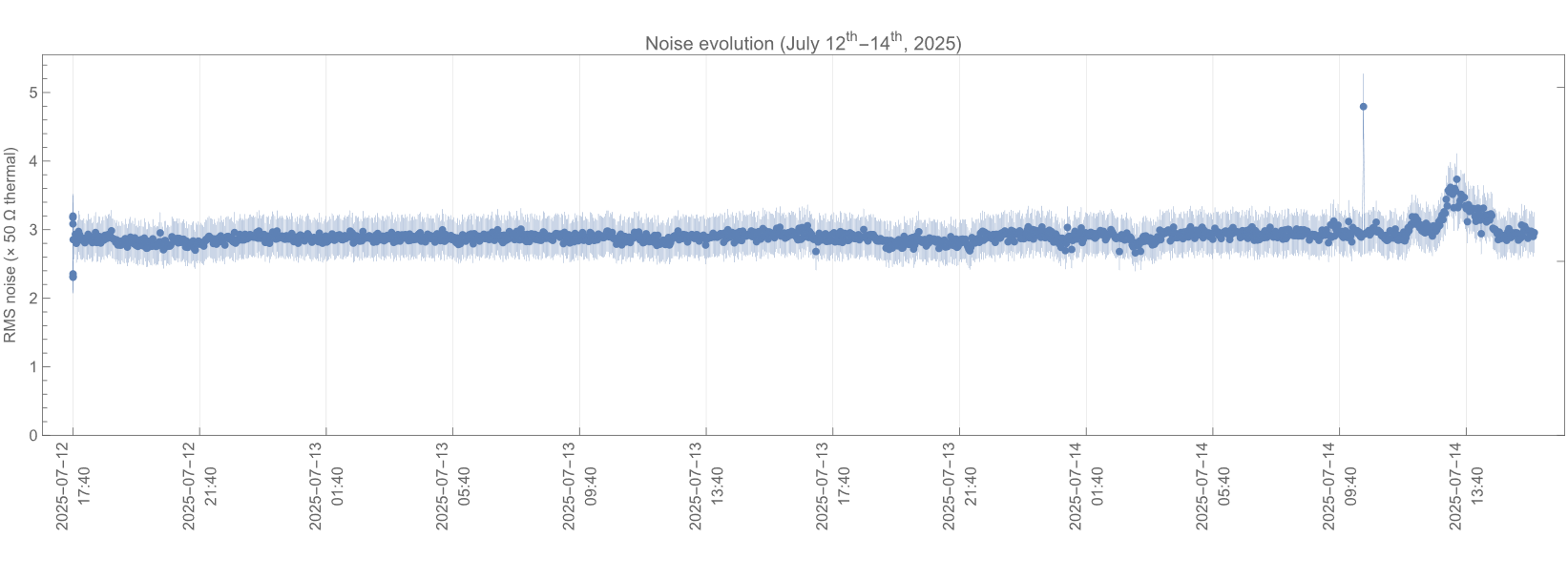}
  \caption{Long-term noise monitoring with the revamped MUST scanner optimally shielded. The plot shows the evolution of the RMS noise level over a continuous period of nearly three days.}
  \label{fig:long_noise_mmt2}
\end{figure*}

Figure~\ref{fig:long_noise_mmt2} shows the RMS noise level monitored continuously over a period of nearly three days. The measurement was carried out with the scanner in its revamped configuration and after proper grounding of all elements.

\subsection{In-vivo imaging}
\begin{figure*}
  \centering
  \includegraphics[width=0.75\textwidth]{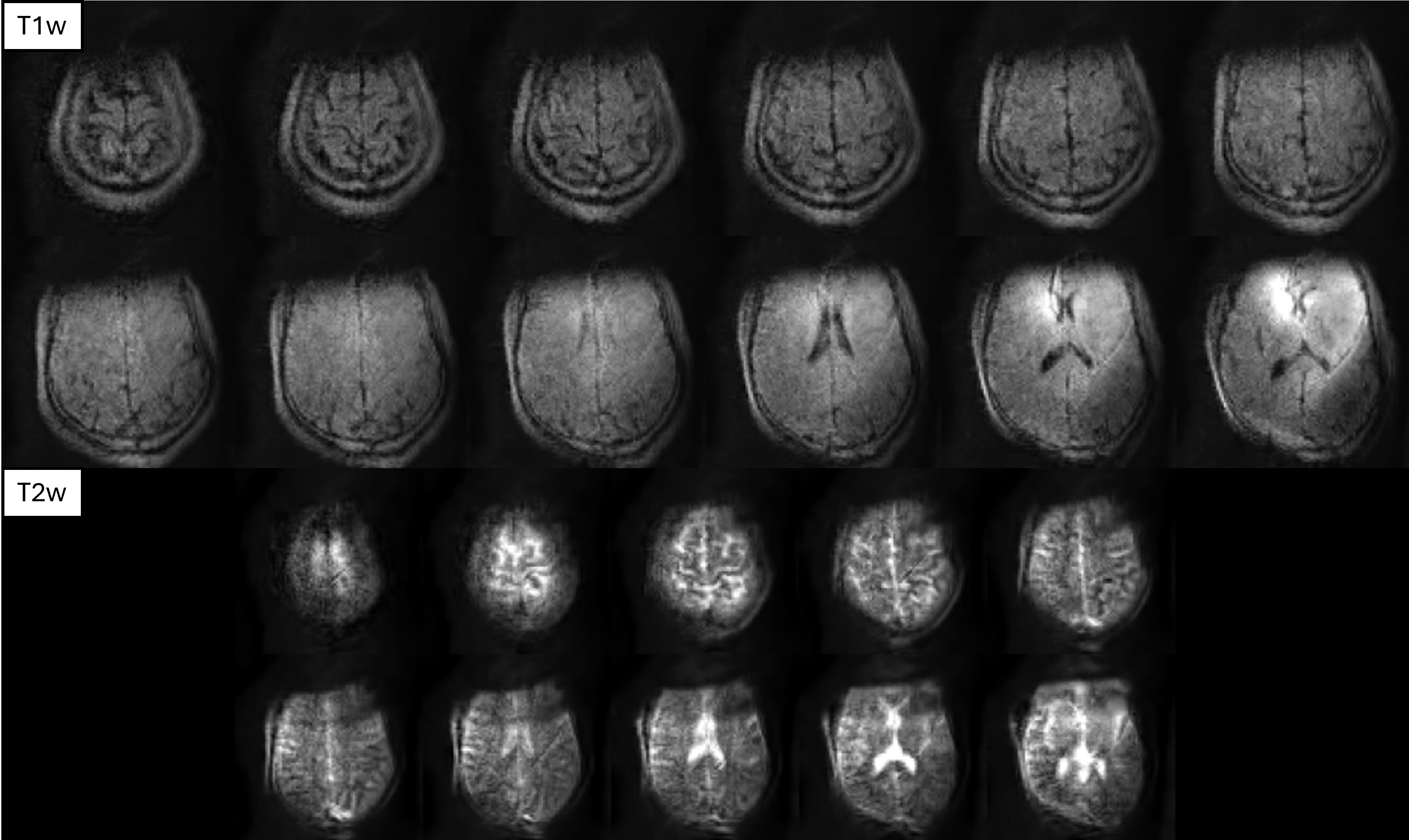}
  \caption{Selected slices from two 3D brain acquisition, a T$_1$-weighted image (top), and a T$_2$-weighted image (bottom). Both datasets were acquired using the updated hardware as described in Sec.~\ref{sec:met_images}. Images were acquired using the parameters listed in Table~\ref{tab:seq_params} and distortion-corrected via Tyger.}
  \label{fig:T1T2}
\end{figure*}

\begin{figure}
  \centering
  \includegraphics[width=\columnwidth]{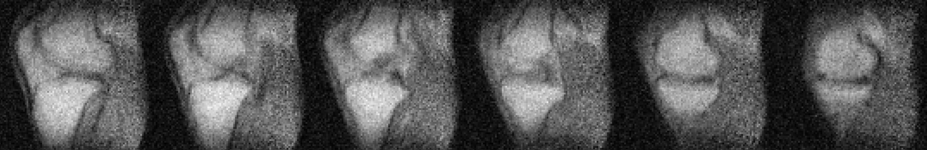}
  \caption{Selected slices from a three-dimensional T$_1$w knee acquisition. Dataset acquired using the updated hardware as described in Sec.~\ref{sec:met_images}, with parameters in Table~\ref{tab:seq_params}.}
  \label{fig:knee}
\end{figure}

\begin{figure*}
  \centering
  \includegraphics[width=\textwidth]{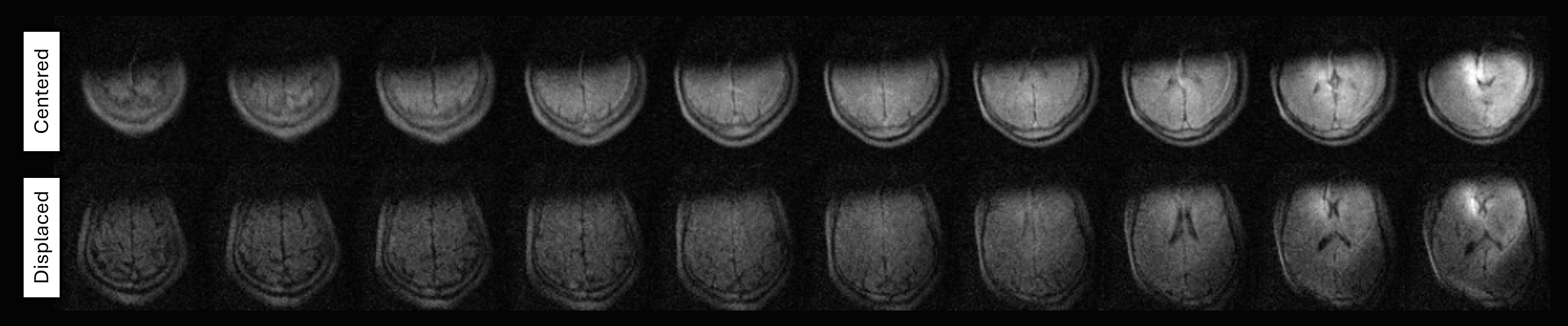}
    \caption{Effect of subject positioning on brain reconstructions. Top row: the subject positioned at the geometric center of the magnet, where $B_0$ inhomogeneity is more pronounced. Bottom row: the same acquisition with the head displaced downwards, toward a region of improved field homogeneity. Images were acquired using the parameters listed in Table~\ref{tab:seq_params} and distortion-corrected via Tyger.}
  \label{fig:displaced}
\end{figure*}

Representative brain and knee images acquired with the upgraded system and reconstructed via Tyger are shown in Fig.~\ref{fig:T1T2} and \ref{fig:knee}.

The influence of subject positioning on image quality is illustrated in Fig.~\ref{fig:displaced}. The top row corresponds to a T$_1$-weighted acquisition with the head positioned at the geometric center of the magnet. The bottom row shows the same subject displaced downwards (closer to the room floor).

\begin{figure}
  \centering
  \includegraphics[width=1\columnwidth]{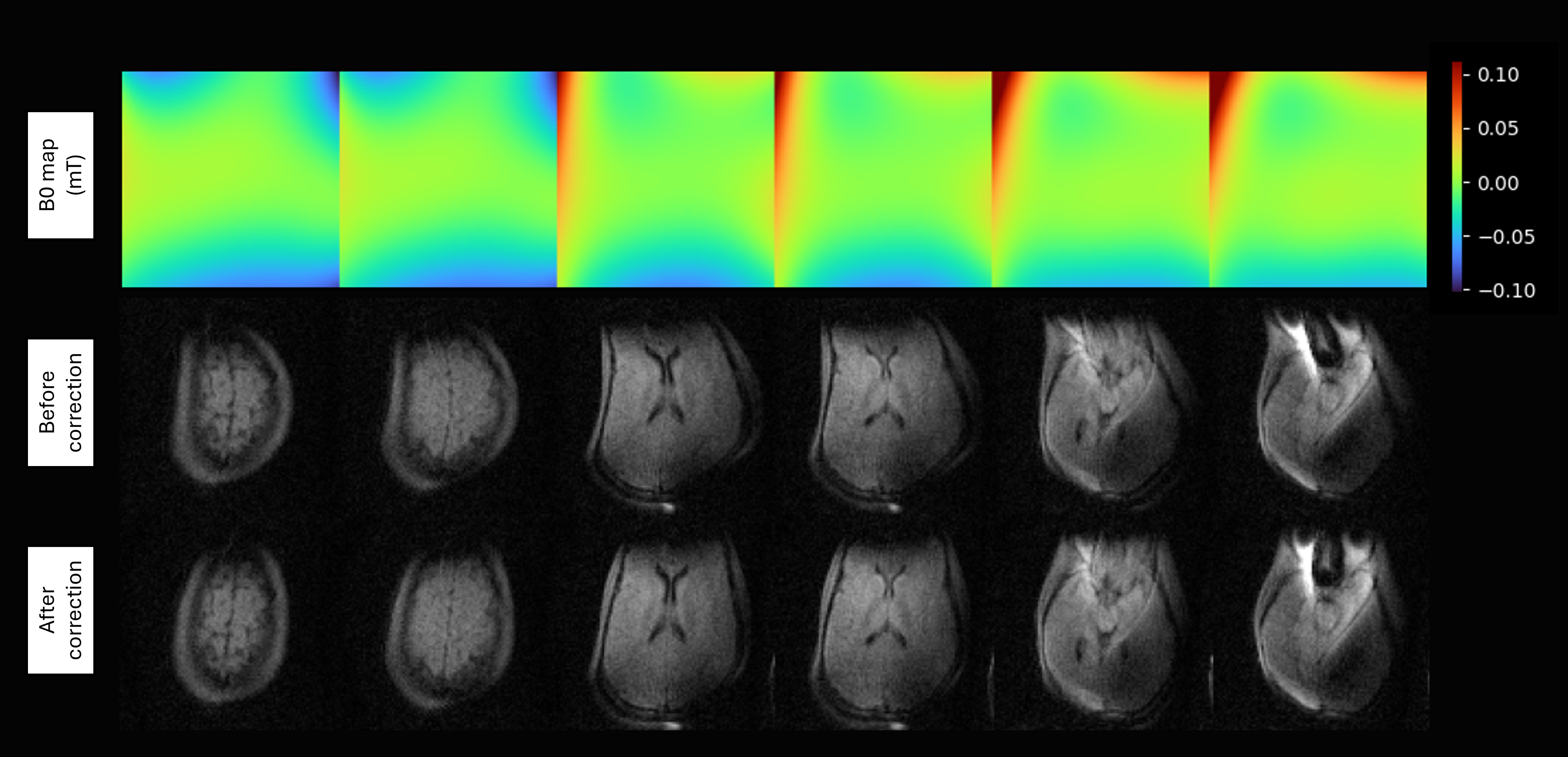}
    \caption{Distortion correction using the SPDS method and cloud-based reconstruction with Tyger. Top row: measured $B_0$ field maps obtained with SPDS~\cite{Borreguero2025}. Middle row: conventional reconstruction without accounting for $B_0$ inhomogeneity. Bottom row: iterative reconstruction incorporating prior knowledge of the measured field map. The correction substantially reduces geometric distortions, particularly in the frontal regions of the brain. Images were acquired using the parameters listed in Table~\ref{tab:seq_params} and distortion-corrected via Tyger.
}
  \label{fig:dist_corr}
\end{figure}

Finally, the effect of distortion correction using SPDS-based iterative reconstruction is presented in Fig.~\ref{fig:dist_corr}. The top row displays the measured $B_0$ field maps, the middle row shows conventional Fourier reconstructions, and the bottom row presents the corrected images.

\section{Discussion}

\subsection{Scanner revamp and electromagnetic noise suppression}
\label{sec:disc_noise}

The noise characterization in Fig.~\ref{fig:protocol}b) provides a quantitative view of the scanner's deviation from the theoretical thermal limit. Even in the simplest configuration (LNA connected to a 50\,$\Omega$ load) the measured noise is about $1.9\times$ the thermal level, indicating that the commercial LNA used in the system introduces a significant contribution. This observation connects directly to the hardware limitations discussed in Sec.~\ref{sec:baseline}: in resource-constrained settings, access to high-performance amplifiers with sub-dB noise figures is limited.

Replacing the load with the RF coil leads to a clear increase in noise, consistent with the coil's intrinsic sensitivity and the coupling of external EMI. Tests performed while powering the system from an uninterruptible power supply (UPS) showed that disconnecting from the mains reduced the noise to near the $2\times$ baseline, confirming the grid as a major noise source. However, the available UPS provided an autonomy of only $\sim$2\,min, precluding continuous battery-powered operation. The highest level ($5.1\times$) was observed when the coil and LNA were connected without proper grounding. As additional grounded elements were incorporated into the setup, the system became progressively more stable.

The long-term noise measurement in Fig.~\ref{fig:long_noise_mmt2} confirms a substantial improvement in electromagnetic stability (with respect to Fig.~\ref{fig:long_noise_mmt}a)) after the complete scanner revamp. Over nearly three days of continuous monitoring, the noise remained close to the achieved floor with minimal drift. Although much of this measurement was acquired during a weekend, the results indicate that the revised grounding and shielding layout effectively mitigated the intermittent noise fluctuations previously observed. In practice, the largest improvements originated from two specific changes: 
(i) establishing a single, low-impedance ground reference for all subsystems, and 
(ii) enclosing the analog electronics in continuous, well-bonded conductive housings. 
The first step eliminated floating reference potentials and reduced common-mode pickup, which had previously led to large fluctuations in the measured noise floor. 
The second minimized radiative coupling into the receive chain, particularly from nearby digital electronics and the power grid. These modifications explain the marked reduction in intermittent noise excursions seen after the revamp. Noise factors below $3\times$ can now be routinely achieved during \emph{in-vivo} operation. Although not ideal, this level suffices for quality data acquisition. The brief increase in noise observed around 13:30 on July~14$^\text{th}$, and a small outlier earlier that same day, are both minor compared with the large fluctuations seen before the upgrade.

All in all, the systematic application of the grounding and shielding guidelines in Ref.~\cite{Guallart-Naval2025a} was critical to achieving stable operation closer to the thermal noise limit.

\subsection{In-vivo imaging}

The \emph{in-vivo} brain images presented in Fig.~\ref{fig:T1T2} represent a significant milestone for the MUST scanner and a clear step forward from the preliminary results of 2023 (Fig.~\ref{fig:setup}b)). While the images are not yet free from imperfections, the leap in performance is evident: anatomical structures are now discernible, the achieved spatial resolution is adequate for low-field operation, and two fundamental contrasts (T$_1$- and T$_2$-weighting) have been successfully demonstrated. These results confirm that the revamped setup, together with the implemented noise suppression strategies and improved control electronics, provides sufficient stability and sensitivity for reliable \emph{in-vivo} imaging.

Signal nonuniformity remains visible, particularly in the upper portions of the images, which correspond to the frontal regions of the head. This darkening arises primarily from the limited output power of the 1\,W RF amplifier, which forced the use of relatively long pulses (around \SI{500}{\micro s}, corresponding to an excitation bandwidth in the order of 2\,kHz). At this level of $B_0$ inhomogeneity (about 4,700\,ppm at 46\,mT, corresponding to a spectral spread of approximately 9\,kHz), the available $B_1$ field cannot excite or refocus the full frequency range present in the unshimmed magnet, so only part of the head is effectively excited. The effect becomes more pronounced when the subject is positioned at the geometric center of the magnet, where $B_0$ is more inhomogeneous (Fig.~\ref{fig:displaced}). This can be alleviated by shifting the head slightly downward, toward regions of improved field uniformity.

Magnetic field inhomogeneity remains the main factor limiting image fidelity. Conventional Fourier reconstruction methods are unable to compensate for the geometric distortions introduced by such strong $B_0$ variations. However, iterative reconstructions that incorporate prior knowledge of the measured field distribution, as obtained with the SPDS mapping method~\cite{Borreguero2025}, substantially improve the resulting images (Fig.~\ref{fig:dist_corr}). Although not eliminating distortions entirely, this approach restores anatomical proportions in the most affected regions, particularly at the periphery. It should be noted, however, that no reconstruction method can recover signal in regions that were not effectively excited in the first place. Because these algorithms are computationally intensive, their integration with Tyger for cloud-based GPU execution enables practical deployment even when local computational resources are limited. The ability to stream reconstructed data back to the console in near real time through \mg{} provides an effective workflow for low-cost systems in low-resource environments. 

Overall, these results demonstrate that, through careful hardware optimization, electromagnetic noise control, and advanced reconstruction pipelines, \emph{in-vivo} imaging is technically achievable on a scanner built and operated in Africa. Although the acquisitions presented here required relatively long scan times, this limitation arises primarily from the elevated noise level of the receive chain, approximately three times the Johnson noise limit. Under ideal conditions, with a lower-noise amplifier and complete isolation from grid interference, comparable image quality could in principle be achieved in roughly one-ninth of the current acquisition time.

\section{Conclusions \& Outlook}
We have presented the first \textit{in-vivo} images acquired with a low-cost MRI scanner built in Africa. This work highlights both the technological feasibility and the capacity-building aspects of deploying affordable MRI systems in low-resource settings, paving the way toward clinical translation.

\begin{figure}
  \centering
  \includegraphics[width=1\columnwidth]{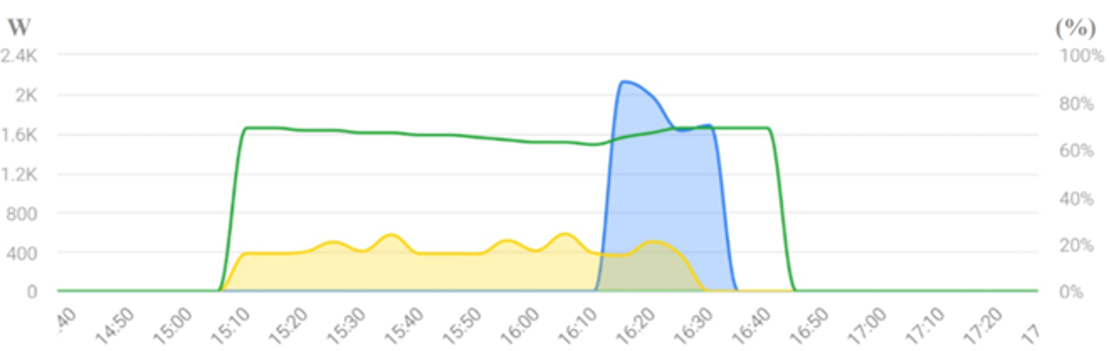}
    \caption{Power consumption and battery state during operation of the 90\,mT NextMRI scanner powered by a BLUETTI AC300 + 2xB300K system. Yellow: instantaneous power drawn by the scanner. Blue: battery charging state. Green: battery charge level.}
  \label{fig:batteries}
\end{figure}

From a technical standpoint, scanner performance is still largely constrained by electromagnetic noise and the instability of power grids in low-resource environments. Recent work with battery-powered systems demonstrates that autonomous operation can overcome many of these limitations. Using a commercial system (BLUETTI AC300 + 2xB300K) as the sole power source for a 90\,mT NextMRI scanner \cite{Galve2024}, stable operation was achieved with power consumption below 600\,W during scanning and approximately 400\,W when idle (Fig.~\ref{fig:batteries}). This configuration provides about 9~hours of continuous MRI operation and allows hot-swappable, seamless recharging while isolating the scanner from mains noise. Extending this approach to autonomous low-field scanners could be relevant to achieving consistent image quality and reliable operation in low-resource settings.

In addition to electromagnetic noise, the other major limitation that currently constrains image quality is the intrinsic $B_0$ inhomogeneity of permanent-magnet Halbach arrays. Distortions arising from inhomogeneities spanning several kHz ultimately limit excitation coverage. While recent open-source initiatives have begun to make field-mapping hardware more accessible, e.g. low-cost robotic arms for characterizing Halbach fields~\cite{Han2017}, the practical implementation of passive shimming remains challenging. There is currently no simple, robust, and broadly accessible methodology for shimming permanent-magnet arrays. Existing strategies typically rely on intricate optimization procedures, custom simulation workflows, or highly-precise inserts that are difficult to design and validate even in well-equipped laboratories. Developing a straightforward, replicable shimming approach suitable for both research and low-resource environments therefore remains an important open challenge for the wider low-field MRI community.

Despite the progress demonstrated here, a number of challenges remain before low-cost MRI can reach widespread and sustainable use in SSA. Capacity building continues to be a major bottleneck, both from the user and clinical perspective and from the technical and industrial side. Low-field MRI technologies are beginning to mature globally, yet local expertise in SSA remains limited, hindering system maintenance, operation, and eventual clinical translation. Developing regional training programs and fostering industrial partnerships will be essential to ensure that locally deployed systems can be effectively utilized and supported \cite{Anazodo2023}. Although pilot clinical studies are under way worldwide, translation to routine healthcare remains slow and will require regulatory frameworks and reliable maintenance pathways that are still in their infancy in low- and middle-income countries. For a more detailed discussion, we refer to Ref.~\cite{Webb2023}.

\appendices

\section*{Contributions}
See Table~\ref{tab:contributions}.

\begin{table*}[t]
\centering
\fontsize{6.5}{7.5}\selectfont
\caption{Author contributions. An “x” indicates participation in the corresponding task.}
\label{tab:contributions}
\begin{tabular}{lccccccccccccccccccc}
\toprule
\textbf{Task} & \textbf{TGN} & \textbf{RAs} & \textbf{PT} & \textbf{MAN} & \textbf{LK} & \textbf{LR} & \textbf{MN} & \textbf{LGCS} & \textbf{MFG} & \textbf{LS} & \textbf{JMA} & \textbf{JS} & \textbf{MH} & \textbf{RAm} & \textbf{AW} & \textbf{JH} & \textbf{SS} & \textbf{JO} & \textbf{JA} \\
\midrule
Scanner revamp      & x & x & x & x & x & x & x & x &   & x & x &   &    & x &   &   &   &   & x \\
Data acq.           & x & x & x & x &   &   &   &   &   &   &   &   &    & x &   &   &   &   & x \\
\mg{} coding        & x &   &   &   &   &   &   &   &   &   & x &   &    &   &   &   &   &   &   \\
Tyger dev.          &   &   &   &   &   &   &   &   &   &   &   & x & x  &   &   &   &   &   &   \\
Tyger integr.       & x &   &   &   &   &   &   &   &   &   & x &   &    &   &   &   &   &   &   \\
Battery tests       & x & x & x & x &   &   &   & x &   &   & x &   &    &   &   &   &   &   & x \\
Proj. conception    & x &   &   &   &   &   &   &   &   &   &   &   &    &   & x & x & x & x & x \\
Proj. manage.       &   &   &   &   &   &   &   &   & x &   &   &   &    & x & x & x & x & x & x \\
Figure prod.        & x &   &   &   & x &   &   &   &   &   &   &   &    &   &   &   &   &   & x \\
Paper writing       & x &   &   &   &   &   &   &   &   &   &   &   &    &   &   &   &   &   & x \\
Paper revision      & x & x & x & x & x & x & x & x & x & x & x & x & x  & x & x & x & x & x & x \\
\bottomrule
\end{tabular}
\normalsize
\end{table*}

\section*{Acknowledgment}
We thank Tom O'Reilly for contributing to the first stages of the scanner revamp and for participating in the conception of the project, and Danny de Gans for assistance restoring the functionality of the GPA. This work was supported by Ministerio de Ciencia e Innovación (PID2022-142719OB-C22), the European Innovation Council (NextMRI 101136407), the ISMRM-Gates Knowledge Exchange Program (91484), and US NIH grant 5R01HD085853-1.



\section*{Conflict of interest}
TGN consults for PhysioMRI Tech. JMA and JA are co-founders of PhysioMRI Tech.


\ifCLASSOPTIONcaptionsoff
  \newpage
\fi


\end{document}